# Plasmon Assisted Optical Curtains

**Yanxia Cui[1, 2], Jun Xu[1], Sailing He[2], and Nicholas X. Fang[1*]**

[1] *Department of Mechanical Science and Engineering and Beckman Institute of Advanced Science and Technology, University of Illinois at Urbana-Champaign, Urbana, Illinois 61801, USA*

[2] *Centre for Optical and Electromagnetic Research, State Key Laboratory of Modern Optical Instrumentation, Zhejiang University; Joint Research Centre of Photonics of the Royal Institute of Technology (Sweden) and Zhejiang University, Zijingang Campus, Zhejiang University, China*

*Corresponding author: nicfang@illinois.edu

We predict an optical curtain effect, i.e., formation of a spatially invariant light field as light emerges from a set of periodic metallic nano-objects. The underlying physical mechanism of generation of this unique optical curtain can be explained in both the spatial domain and the wave-vector domain. In particular, in each period we use one metallic nanostrip to equate the amplitudes of lights impinging on the openings of two metallic nanoslits and also shift their phases by $\pi$ difference. We elaborate the influence on the output effect from some geometrical parameters like the periodicity, the slit height and so on. By controlling the light illuminated on metallic subwavelength apertures, it is practical to generate optical curtains of arbitrary forms, which may open new routes of plasmonic nano-lithography.

Recently, rapid development of research has taken place in the area of surface plasmon polaritons, i.e., collective electron excitations supported on metallic nano-objects [1]. Although this phenomenon is well known for decades, the research has gained new momentum in advantage of state-of-the-art nanofabrication techniques [2] and powerful electromagnetic simulation methods [3]. Because of the promise of surface plasmon polaritons to manipulate light far beyond the diffraction limit [4-6], recent efforts are ongoing to miniaturize conventional light emitting devices [7], lasers [8], sensors [9] and photovoltaic cells [10] at unprecedented nanoscale dimensions.

For instance, plasmonic apertures in subwavelength scale have been proposed as core elements of new optical devices that employ plasmon excitations to track single molecule [11], focus the signal towards a detector [12], or change beam shape [13]. One metallic subwavelength aperture could be regarded as a point source in terms of conventional optics as the wave on the output opening of the aperture would diffract into all radial directions, by the coupling between plasmon waves and radiation modes [14, 15]. Theoretical efforts on transmission properties of two dimensional periodic metallic nanoslits have between reported both numerically [16] and semi-analytically [3]. These studies also lead to possible applications in nanolithography in near field [18] and far-field [19]. However, the study on plasmonic apertures is still infancy since most of previous researches are limited to designing structural features milled in a single metallic layer [13-19]. More recently, Wang *et. al* [20] demonstrated a plasmonic multilayer structure consisting of two longitudinally cascaded metallic slit arrays with a half pitch off-set which could be used as a filter. In this paper, we will also focus our study on a multilayer plasmonic structure but from a quite different point of view. By introducing additional metallic nanoobjects above the metallic subwavelength apertures, we will show the control of both the amplitude and phase of light impinging on these apertures. As a result, we obtain a unique optical curtain on the output region of these apertures based on which more exciting applications might be discovered in future.

The sketch of the proposed system under study is shown in Fig. 1. Periodic nanoslit apertures are milled in a thin metal film. These nanoslits under illumination work as a series of line sources, delivering cylindrical waves from their openings. On top of the nanoslits, we add metallic nanostrips to control the amplitude and phase of light illuminating on the input openings of those slits. We define the periodicity of the structure as $P$; in each period, the metallic nanostrip (of width $W_p$ and height $h$) is put with a separation $d$ above two metallic nanoslits (of width $W_s$ and height $t$). In other words, the metallic nanoslits are of periodicity $P/2$. In our study, the nano-system is normally illuminated by a plane wave of TM polarization [the magnetic field ($H_y$) is perpendicular to the *x-z* plane].

For simplicity without loss of generality, in this letter we use a noble metal such as silver at wavelength $\lambda_0 = 1$ μm in vacuum. We assumed the permittivity of silver as $\varepsilon_m = -48.8 + i3.16$ [15, 16]. All the results are calculated by commercial software COMSOL based on finite element method [3]. Perfect matched layers are applied along *z* direction to avoid artificial reflections from the boundaries. Non-uniform mesh sizes are used to ensure the numerical convergence as well as to save the memory. Our numerical method has been validated with the results for the structures studied in Ref. [16].

As shown in Fig 2, a unique optical curtain is observed when light is passing through the proposed plasmonic nano-system. The field amplitude ($|H_y|$, $|E_x|$ and $|E_z|$) has no variation along *z* direction and there are two peaks and two valleys in each period along *x* direction. In order to observe this effect, we set the nanostrips as $W_p = 900$ nm and $h = 300$ nm, and the separation $d = 110$ nm. We set a fixed value of the slit width ($W_s$) of 50 nm and the slit height ($t$) of 460 nm so the bare nanoslit is non-resonant [in a fashion similar to a vertical metal-insulator-metal (MIM) waveguide] [6, 15]. We consider the periodicity ($P$) in the range of ($\lambda_0$, $2\lambda_0$) and first of all assume it as 1100 nm.

In contrast, when only a metallic nanoslit array with periodicity $P$ is illuminated (i.e., removing both the above nanostrips and the middle nanoslits opened right under the nanostrips) as shown in Fig. 3(a-c), some oscillations are produced along *z* axis for $|H_y|$ and $|E_x|$, but there is still no oscillations for $|E_z|$. Such unique property of periodic metallic nanoapertures are reported by Shao and Chen [19] for direct patterning of three-dimensional periodic nanostructures. Our work could contribute to making this problem explicit.

The results obtained in Fig. 2 and Fig. 3(a-c) reveals the importance of added open nanoslits at positions $x = NP$ ($N$ is an arbitrary integer). For the structure in Fig. 2, we observed a $\pi$ phase difference between the magnetic field at the openings of neighboring nanoslits spaced by $P/2$, as plotted in Fig. 2(d). This $\pi$ phase difference is due to the retarded field when the scattered wave propagates horizontally in the gap between the two metallic layers [21-22]; similar effects are also reported in vertical MIM waveguide, such as in Ref. [15]. In Fig. 2, we also notice that the amplitude of the energy emitted from the output openings of the nanoslits at positions $x = NP$ are almost as same as that at positions $x = (2N+1) P/2$. The channel of the nanoslit is to deliver the energy from its input to its output. Thus, it is intuitive that the amplitude of light impinging on the input openings of all the nanoslits are identical. On the other hand, if we only open additional nanoslits at positions $x = NP$, without the double metallic layer, the neighboring nanoslits of periodicity $P/2$ are of same phase, smearing out the interference effect needed for the optical curtains. Additionally, it is important to emphasize that for both structures in Fig. 2 and Fig. 3(a-c), the perpendicular component of the electric filed ($|E_z|$) has the distribution without any variation along $z$ direction. This will be explained later in this paper.

In the following, we will simplify the proposed plasmonic structure in Fig. 2 to obtain its physical model in the spatial domain. The metallic nanoslit could be considered as a line source to generate a cylindrical wave, that can be written as $H_y(x,z) = A \frac{1}{\sqrt{|\vec{r}|}} e^{ik_0|\vec{r}|+i\phi}$, with $\vec{r}[x_s,z_s] = (x-x_s)\hat{x} + (z-z_s)\hat{z}$, where $[x_s,z_s]$ are the coordinates of the slit's output opening (center position), while $A$ and $\phi$ are its amplitude and phase, respectively. For our well-designed system in Fig. 2, the amplitude and phase of the cylindrical wave are exactly determined by the incident condition at the input openings of the metallic nanoslits. Based on our previous analysis, we could replace the two metallic nanoslits in each period by two cylindrical waves which have the same amplitude but a $\pi$ phase difference that are shown in Fig. 4(a). Therefore, the total field distribution could be calculated by a superposition of cylindrical waves according to the following equation:

$$H_y(x,z) = \sum_{N=-\infty}^{\infty} \frac{1}{\sqrt{|\vec{r}[NP,0]|}} e^{ik_0|\vec{r}[NP,0]|} + \sum_{N=-\infty}^{\infty} \frac{1}{\sqrt{|\vec{r}[\frac{2N+1}{2}P,0]|}} e^{ik_0|\vec{r}[\frac{2N+1}{2}P,0]|+i\pi}$$

Thus, we calculate the field distribution and plot it in Fig. 4(c), which shows excellent agreement with the result in Fig. 2(a) calculated in COMSOL. By modifying the equation to $\sum_{N=-\infty}^{\infty} \frac{1}{\sqrt{\left|\vec{r}[\frac{2N+1}{2}P,0]\right|}} e^{ik_0\left|\vec{r}[\frac{2N+1}{2}P,0]\right|}$, we could obtain the field distribution for the structure in Fig. 3(a-c) which is also in agreement with the result calculated in COMSOL; this confirms the three dimensional periodical patterns reported in Ref. [19].

Due to the periodic nature of our system, we could model it in the wave-vector domain as well. For the problem in Fig. 2, the total field can be considered as the superposition of two periodic source arrays. One of them is located at $x = NP$ (labeled as Array 1) and the other is located at $x = (2N+1) P/2$ (labeled as Array 2). These two source arrays are off-set with a half period by a $\pi$ phase difference. Thus, the total field distribution could be regarded as a superposition of the effects from the diffractive beams of these two source arrays. These diffracting beams have different propagating directions, described by wave-vectors. The $x$ component of the wave-vector for order $M$ (an integer) has the value $k_{xM} = M\frac{2\pi}{P}$. Therefore the total wave-vector for the order $M$ is $\vec{K}_M = k_{xM}\hat{x} + k_{zM}\hat{z} = k_{xM}\hat{x} + \sqrt{k_0^2 - k_{xM}^2}\hat{z}$, where $k_0 = \frac{2\pi}{\lambda_0}$ represents the wave-vector in the free space of the output domain. Thus, the field distribution could be written as the superposition of a set of plane waves with different diffraction orders $H_y(x,z) = \sum_{M=-l}^{l} e^{i\vec{K}_M \vec{r}[0,0]} + B \sum_{M=-l}^{l} e^{i\vec{K}_M \vec{r}[P/2,0] + i\theta}$, in which $B$ and $\theta$ represent the relative amplitude and phase of Source Array 2 in comparison with Source Array 1, and $l$ is the maximum diffracting order allowed to propagate into far field, which is determined by the periodicity. For the periodicity in the range ($\lambda_0$, $2\lambda_0$), $l$ equals to 1 because only the diffraction modes of 0th and ±1st orders can propagate along $z$ direction, and higher order modes are evanescent. Apparently for the special case in Fig. 2, $B = 1$ and $\theta = \pi$. Hence, we could draw in Fig. 4(b) the physical model in the wave-vector domain for our proposed structure. Again, the calculated result agrees with the result computed in COMSOL. Furthermore, one can find out that, for Source Array 1 and 2, their 0th orders interfere destructively along $z$ direction because of their $\pi$ phase shift, while the ±1st orders are of the same form due to an additional $P/2$ position off-set along $x$ direction. Because of that, the model for the system in Fig. 2 could be further simplified as the superposition of two diffracting plane waves (±1st orders) and finally written as $H_y(x,z) = e^{i\vec{K}_1 \vec{r}[0,0]} + e^{i\vec{K}_{-1}\vec{r}[0,0]}$.

Based on the above physical model in the wave-vector domain, we can also confirm why the patterns for the $|E_z|$ component in Fig. 2(c) and Fig. 3(c) have no variations along the perpendicular direction. Considering the situation in Fig. 3(c) as an example,

based on the equation $E_z \propto \frac{\partial H_y}{\partial x}$ for TM illumination, $E_z$ component could be written as $E_z(x,z) \propto \sum_{M=-l}^{l} ik_{xM} e^{i\vec{K}_M \vec{r}[0,0]}$. Then, we could get its differential coefficient formula along $z$ direction as $\frac{dE_z(x,z)}{dz} \propto \sum_{M=-l}^{l} -M\frac{2\pi}{P}\sqrt{k_0^2 - (M\frac{2\pi}{P})^2} e^{i\vec{K}_M \vec{r}[0,0]}$. It is straightforward to achieve that the 0th order has no contribution and the $\pm M$ th orders are cancelled each other. The same result could be obtained for the situation in Fig. 2(c), too.

For the interest of nanolithography applications, it is noteworthy to estimate the contrast of the output pattern for local light intensity ($I = EE^* = E_x^2 + E_z^2$) of the proposed plasmonic nano-system. At a cross plane of constant $z$, the contrast of the pattern is defined as $c = \frac{I_{max} - I_{min}}{I_{max} + I_{min}}$. In Fig. 5(a), we show the relation between the contrast ($c$) and the period [$P$, in the range ($\lambda_0$, $2\lambda_0$)] which can be obtained by simulating in COMSOL or calculating based on the physical models. Note that the wave-vector domain model could directly produce the analytical formula as $c = \left|\frac{2\lambda^2}{P^2} - 1\right|$ according to equations ($E_x \propto H_y$ and $E_z \propto \frac{\partial H_y}{\partial x}$). It is observed that the contrast approaches zero when the period equals 1414 nm. This is because along $x$ direction, the $|E_z|$ component is maximal exactly at the zero point position where the $|E_x|$ component is minimal; and vice versa. Besides, for different periodicities, the electric field is decomposed into its two components which have different proportions, thus the superposition of them will give different intensity contrasts for the total electric field. It is no doubt that at the particular case $P$ = 1414 nm, the amplitudes of the two electric field components are same so that the field distribution on the output side of the apertures is uniform.

Additionally, we tune the parameter of the slit height ($t$) and show in Fig. 5(b) its relationship with the maximum and minimum values of the amplitude of the electric field ($|E|$) along a cross-sectional plane at distance $z$ = 2 μm. The curves oscillate with increasing slit height, showing a clear Fabry-Perot (F-P) effect in the single metallic nanoslit [16, 21]. Fig. 5(c) show the influence of the strip height ($h$). We find the electric field suffers only some tiny oscillations, which means that the strip height has very weak influence on this optical curtain effect. Actually, the same F-P effect happens. But at this time, the gaps between neighboring nanostrips are much wider than the width of the nanoslits, thus the energy passing through the gaps between nanostrips would not vibrate as strongly as that shown in Fig. 5(b). Then, we calculate the contrast ($c$) with varied slit height (or strip height) and show them in the inset of

Fig. 5(b) and (c). We see that when *t* (or *h*) is larger than 100 nm, the contrast is almost around the value 0.653 which could be calculated from $c = \left| \frac{2\lambda^2}{P^2} - 1 \right|$. For the situations when *t* (or *h*) is less than 100 nm, there will be some light directly penetrating through the metallic film, thus the generated pattern would have some oscillating fringes along the perpendicular direction and looks different from that shown in Fig. 2 based on which we define the contrast. Thus, the contrast is no longer invariant along the propagation direction, when the slit height (or strip height) is less than 100 nm. Besides, when the metallic nanoslits are resonant (e.g., *t* = 290 nm), the contrast deviates slightly from 0.653. This is due to the strong impact of the resonant nanoslits on the incident light impinging on the openings of the nanoslits [21-22].. When the incident light on neighboring nanoslits is not in conformity with that in Fig. 2 (i.e., the same amplitude and $\pi$ different phase), the formed pattern in the output region would suffer some oscillation along the *z* direction, too. Fortunately, this divergence is quite small and negligible for practical applications.

According to Fig. 5(b) and (c), we could deduce that an optical curtain can be observed on the nanoslits and the nanostrips as thin as 100 nm. This might help to save precious noble metals in practical experiments. Also note that in order to keep the output distribution identical to that in Fig. 2, other parameters should also be considered. For example, if we set the parameters in Fig. 2 by shortening $W_p$ to 400 nm and *d* to 110 nm with other parameters unaltered, some oscillating fringes would show up along *z* direction as shown in Fig. 3(d) (only $|H_y|$ field is illustrated). This is because the requirements for light illumination on the openings of neighboring nanoslits (i.e., the same amplitude and $\pi$ different phase) can no longer be satisfied.

In conclusion, we have proposed a novel plasmonic nano-system by putting a metallic nanostrip array above a periodic metallic nanoslit array to generate a special optical curtain which has no field variation along the perpendicular direction. The optical curtain is the result of precise control of phase and amplitude on the neighboring metallic subwavelength apertures. We have linked our work with conventional optics and explained it physically in both the spatial and wave-vector domains. Aiming at forming an optical curtain invariant along the perpendicular direction, the structural parameters of the above nanostrips should be carefully chosen so that both amplitude and phase of the illuminated light on the apertures could fulfill the required conditions. We emphasize that the plasmonic apertures (i.e., the slits in 2-D space) should be in subwavelength scale to make sure that they could be regarded as point sources (or line sources in 2-D space).

The approaches outlined in our work would be helpful to design and predict the optical field pattern formation by plasmonic aperture arrays [19], that are particularly interesting for nanolithography techniques. Furthermore, the simplicity of our analysis could lend itself to the design of more complex optical curtains which might be produced in future, since we may be able to freely control the phase and amplitude of individual plasmonic apertures for the formation of arbitrary patterns. Both the experimental verification and design of three-dimensional devices are in progress.

This work is partially supported by the National Science Foundation (CMMI 0846771) and the National Science Foundation of China (60688401). Yanxia Cui would like to thank Dr. Kin Hung Fung for helpful discussions.

# Figure Captions

Fig. 1. (Color online) Sketch of the proposed periodic metallic nano-system composed of metallic nanostrip array and metallic nanoslit array. The metallic nanostrips are of width $W_p$, height $h$ and periodicity $P$, and the metallic nanoslits are of width $W_s$, height $t$ and periodicity $P/2$. A plane wave of TM polarization is illuminating with wavelength $\lambda_0 = 1$ $\mu$m at normal incidence. The metallic layers are in green color.

Fig. 2. (Color online) Distributions of field [$|H_y|$, $|E_x|$, $|E_z|$ and $\Phi(H_y)$] for the proposed system. $W_s = 50$ nm, $t = 460$ nm, $P = 1100$ nm, $W_p = 900$ nm, $h = 300$ nm, $d = 110$ nm. The red arrows in the output region in (a) represent the poynting vectors.

Fig. 3. (Color online) (a-c) Distributions of field ($|H_y|$, $|E_x|$ and $|E_z|$) for metallic nanoslits array of periodicity $P$. (d) Distributions of field $|H_y|$ by tuning the strip width in Fig. 2 to 400 nm and the separation to 110 nm. The red arrows in the output region in (a) and (d) represent the poynting vectors.

Fig. 4. (Color online) (a) The physical model in the spatial domain of the problem described in Fig. 1. (b) The model in the wave-vector domain when the periodicity is in the range of ($\lambda_0$, $2\lambda_0$). (c) The calculated distribution of magnetic field ($|H_y|$) based on the physical model in the spatial domain for the structure in Fig. 2.

Fig. 5. (a) Relationship of the intensity contrast ($c$) and periodicity ($P$) at $\lambda_0 = 1$ $\mu$m. (b-c) Relations between maximum and minimum electric fields in the plane $z = 2$ $\mu$m (labeled as $|E_{max}|$ and $|E_{min}|$) and the nanoslit height $t$ or the nanostrip height $h$, insets: Relations between the intensity contrast ($c$) and the slit height (or the strip height).

Fig. 1. Cui, Xu, He and Fang

Fig. 2. Cui, Xu, He and Fang

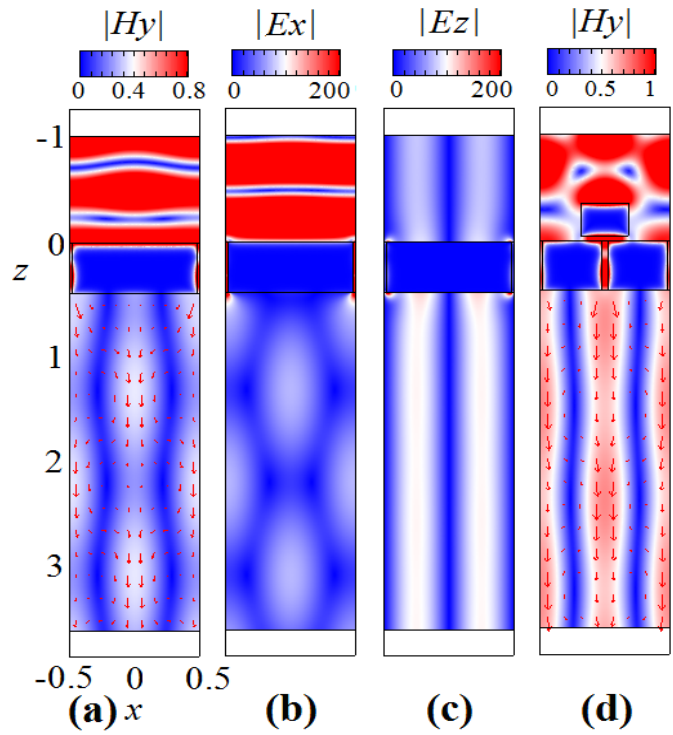

Fig. 3. Cui, Xu, He and Fang

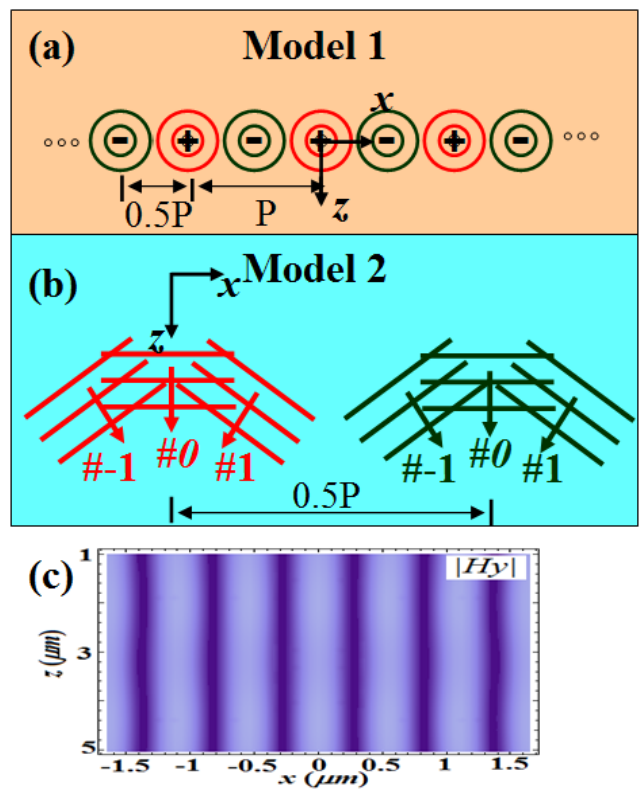

Fig. 4. Cui, Xu, He and Fang

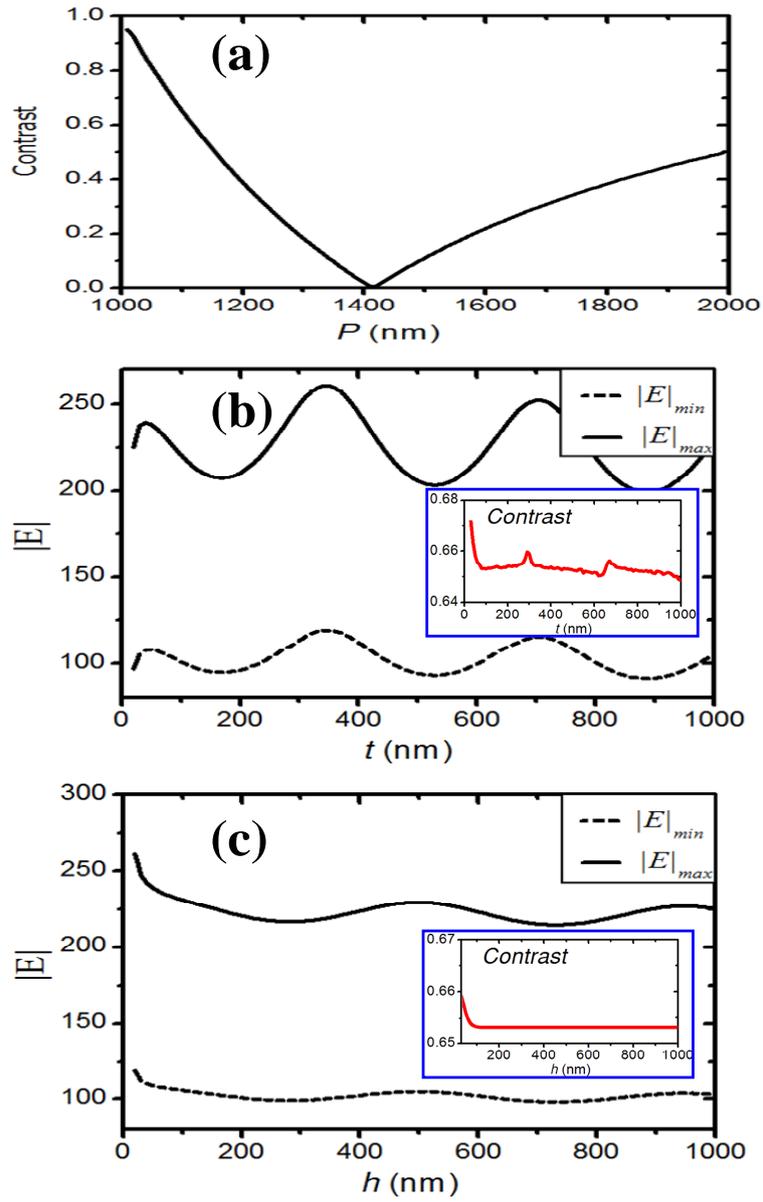

Fig. 5. Cui, Xu, He and Fang